\documentclass[conference]{IEEEtran}

\makeatletter
\def\ps@headings{%
\def\@oddhead{\mbox{}\scriptsize\rightmark \hfil \thepage}%
\def\@evenhead{\scriptsize\thepage \hfil \leftmark\mbox{}}%
\def\@oddfoot{}%
\def\@evenfoot{}}
\makeatother
\usepackage[latin1]{inputenc}
\usepackage{tikz}
\usetikzlibrary{shapes,arrows}
\usepackage{graphicx,cite}
\usepackage{multirow}
\usepackage{epsfig}
\usepackage[cmex10]{amsmath}
\usepackage{array}

\usepackage{amssymb}
\usepackage{array}
\usepackage{fixltx2e}
\usepackage{amsfonts}

\begin{document}
\title{Wireless Backhaul Node Placement for Small Cell Networks}
\author{\IEEEauthorblockN{Muhammad Nazmul Islam\IEEEauthorrefmark{0}, 
Ashwin Sampath\IEEEauthorrefmark{1} , Atul Maharshi\IEEEauthorrefmark{1},
Ozge Koymen\IEEEauthorrefmark{1} and
Narayan B. Mandayam\IEEEauthorrefmark{0} \\ }
 \IEEEauthorblockA{\IEEEauthorrefmark{0}
%Dept.~of Elec.~and Comp.~Eng.,
WINLAB, Rutgers University, 
Email: \{mnislam,narayan\}@winlab.rutgers.edu \\
%Email: mnislam@winlab.rutgers.edu, narayan@winlab.rutgers.edu \\
\IEEEauthorrefmark{1} Qualcomm Corporate R\&D, 
Email: \{asampath,atulm,okoymen\}@qti.qualcomm.com}}

\maketitle

\begin{abstract}

Small cells have been proposed as a vehicle for wireless networks 
to keep up with surging demand. Small cells come with a significant challenge of 
providing backhaul to transport data to(from) a gateway 
node in the core network. Fiber based backhaul offers the high rates 
needed to meet this requirement, but is costly and time-consuming to 
deploy, when not readily available. Wireless backhaul 
is an attractive option for small cells as it provides a less expensive 
and easy-to-deploy alternative to fiber. However, there are multitude 
of bands and features (e.g. LOS/NLOS, spatial multiplexing 
etc.) associated with wireless backhaul that need to be used intelligently for 
small cells.  Candidate bands include: sub-6 GHz band
that is useful in non-line-of-sight (NLOS)
scenarios, microwave band ($6-42$ GHz) that is useful in point-to-point 
line-of-sight (LOS) scenarios, and millimeter wave bands (e.g. $60$, $70$ and $80$ GHz) 
that are recently being commercially used in LOS scenarios.  In many deployment topologies, it is advantageous to use aggregator nodes, located at 
the roof tops of tall buildings near small cells. These nodes 
can provide high data rate to multiple small cells in NLOS paths, 
sustain the same data rate to gateway nodes using LOS paths and 
take advantage of all available bands. This work performs the 
joint cost optimal aggregator node placement, power allocation, 
channel scheduling and routing to optimize the wireless backhaul network. 
We formulate mixed integer nonlinear programs (MINLP) to 
capture the different interference and multiplexing patterns at 
sub-6 GHz and microwave band. We solve the MINLP through linear 
relaxation and branch-and-bound algorithm and apply our algorithm 
in an example wireless backhaul network of downtown Manhattan.

\end{abstract}


\section{Introduction}

Demand for wireless services is increasing rapidly. Some industry and academic experts predict a $1000$-fold demand increase
by $2020$~\cite{QualcommChallenge,Andrews1}. Improvements in physical layer alone 
cannot sustain such high data rate~\cite{Andrews1,PhyLayerDead}. Extreme densification
of heterogeneous small cells is necessary to meet this demand~\cite{FemtocellSurvey}.
Small cells come with a significant challenge of 
providing backhaul to transport data to(from) a gateway 
node (node with existing fiber point, often co-located with a macrocell) 
in the core network. Fiber based backhaul offers the high rates 
needed to meet this requirement, but is costly~\cite{Ceragon} and 
time-consuming to deploy, when not readily available.

\begin{figure}[t]
\centering
\includegraphics[scale=0.4]{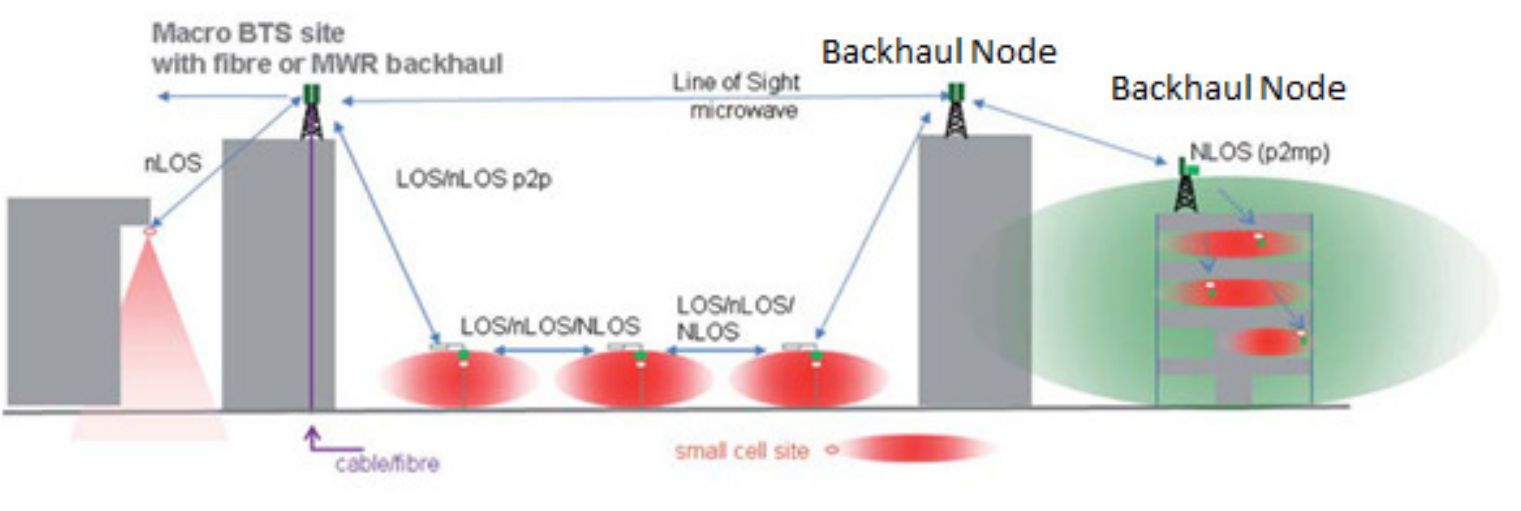}
\caption{
%The figure shows the necessity of backhaul nodes in
%a metropolitan setting.
The macrocell is located at the second
building from the left and it has fiber connection. The red bubbles denote
coverage of the small cells.
The small cells located at the left side can
directly communicate with the macrocell. The small cells located at the
right side do not have good one-hop wireless links
with the macrocell and require routing through backhaul nodes
to reach the macrocell. This figure is reproduced from~\cite{NGMN}}
\centering
\label{fig:Generic_Figure_AN}
\end{figure}

Wireless backhaul can be a valuable option in this regard.
One needs to utilize the available bands judiciously to attain
the maximum advantage of wireless backhaul.  
There are multiple candidate bands for wireless backhaul: 
first, sub-$6$ GHz band that is useful in non-line-of-sight (NLOS) 
point-to-multipoint scenarios,
microwave band ($6-42$ GHz) that is useful in line-of-sight (LOS)
point-to-point scenarios and is currently going through experimental
research in NLOS scenarios~\cite{Ericsson1}, 
and millimetre wave band ($60$, $70$ and $80$ GHz) that is recently
being commercially used in LOS scenarios.
Small cells that are located at lamp posts, street corners, 
low rooftops, etc., may not have line-of-sight (LOS) path to the 
gateway nodes. An effective way to utilize the available bands for wireless backhaul 
is to place aggregator nodes at the tall buildings 
that are located close to small cells. 
Aggregator nodes can provide high data rate to multiple small cells in NLOS paths, 
sustain the same data rate to gateway nodes using LOS paths and 
take advantage of all available bands.
Fig.~\ref{fig:Generic_Figure_AN} shows the importance of
aggregator nodes. 

Deployment of aggregator nodes in roof tops of tall 
buildings consume operational leasing cost.
The network operator needs to minimize the operational expenses
while ensuring the network connectivity between the
small cell and the gateway node.
Hence, optimal aggregator node deployment and network
connectivity design to minimize the operational
expenses is essential for small cell networks.

%Optimization of wireless backhaul networks for small cells
%depends on the selected spectrum for different bands.
%There are couple of candidate frequencies that one 
%can use between small cells and aggregator
%nodes: first, sub-6 GHz band (mainly 5.8 GHz band) and second, microwave band (mainly 28 GHz band). 
%interference pattern, i.e., interference between two links that are not adjacent to
%each other, differs in these two scenarios.
%This work reviews network optimization problems in both these cases. 
%While sub-$6$ GHz based commercial radios already operate in non-line-of-sight (NLOS) %paths,
%the operation of microwave band in NLOS direction is still going through experimental
%research~\cite{Ericsson1}. 

%Sub-6 GHz spectrum allows aggregator nodes to use large multiple-input multiple-output (MIMO) technology~\cite{LargeMIMO}
%and communicate with multiple small cells at the same time and frequency slot. 
%Due to the narrow beam width of its antennas, $28$ GHz band 
%allows high antenna gain and low interference regime and it may
%lead to higher throughput even in non-line-of-sight (NLOS) scenarios~\cite{Ericsson1}.
%Network optimization problem changes from interference limited regime (5.8 GHz)
%to interference free regime (60 GHz) in these two different scenarios.

We perform joint cost optimal aggregator node placement, power control, 
channel scheduling and routing to minimize operational expenses
of the overall network. We develop a mixed integer non-linear
programming (MINLP) formulation and then convert the MINLP 
to a mixed integer linear program (MILP) using linear relaxation techniques.
We apply the MILP based optimization results
to solve the network connectivity in an example downtown Manhattan scenario.

\subsection{Related Works}

%Lin et. al. and Lu. et. al. have developed relay station placement strategies in
%IEEE 802.16j WiMAX networks in~\cite{RelayPlacement} and~\cite{RelayPlacement2}.
%Their works connect static base stations 
Relay or aggregator node placement problems have appeared 
in different scenarios,
such as wireless sensor networks (WSN)~\cite{Sensor1,Sensor2}, 
wireless local area networks (WLAN)~\cite{WLAN1} 
and IEEE 802.16j WiMAX networks~\cite{WiMAX1,WiMAX2}.

In~\cite{Sensor1}, the authors assume a radius coverage based 
propagation model, and deploy sensor and relay nodes optimally
among an unconstrained number of candidate locations to solve
the connectivity and routing problem. The authors of~\cite{Sensor2} 
deploy sensor and relay nodes among a constrained set of candidate
locations but assume a radius coverage based propagation model.
%In~\cite{Mesh1}, the authors formulate relay
%placement problems in wireless mesh networks as integer programs
%and solve it with Bender's decomposition. 
The authors of~\cite{WLAN1}
solve the relay placement problem in WLAN with uniformly distributed
mobile users. In~\cite{WiMAX2}, the authors solve relay placement
problem in IEEE 802.16j networks while assuming an arbitrary user distribution
and distance based propagation. The authors of~\cite{WiMAX1} also 
focus on IEEE 802.16j networks, assume nomadic relay nodes 
and place relays for time varying user demand.

\begin{figure}[t]
\centering
\includegraphics[scale=0.5]{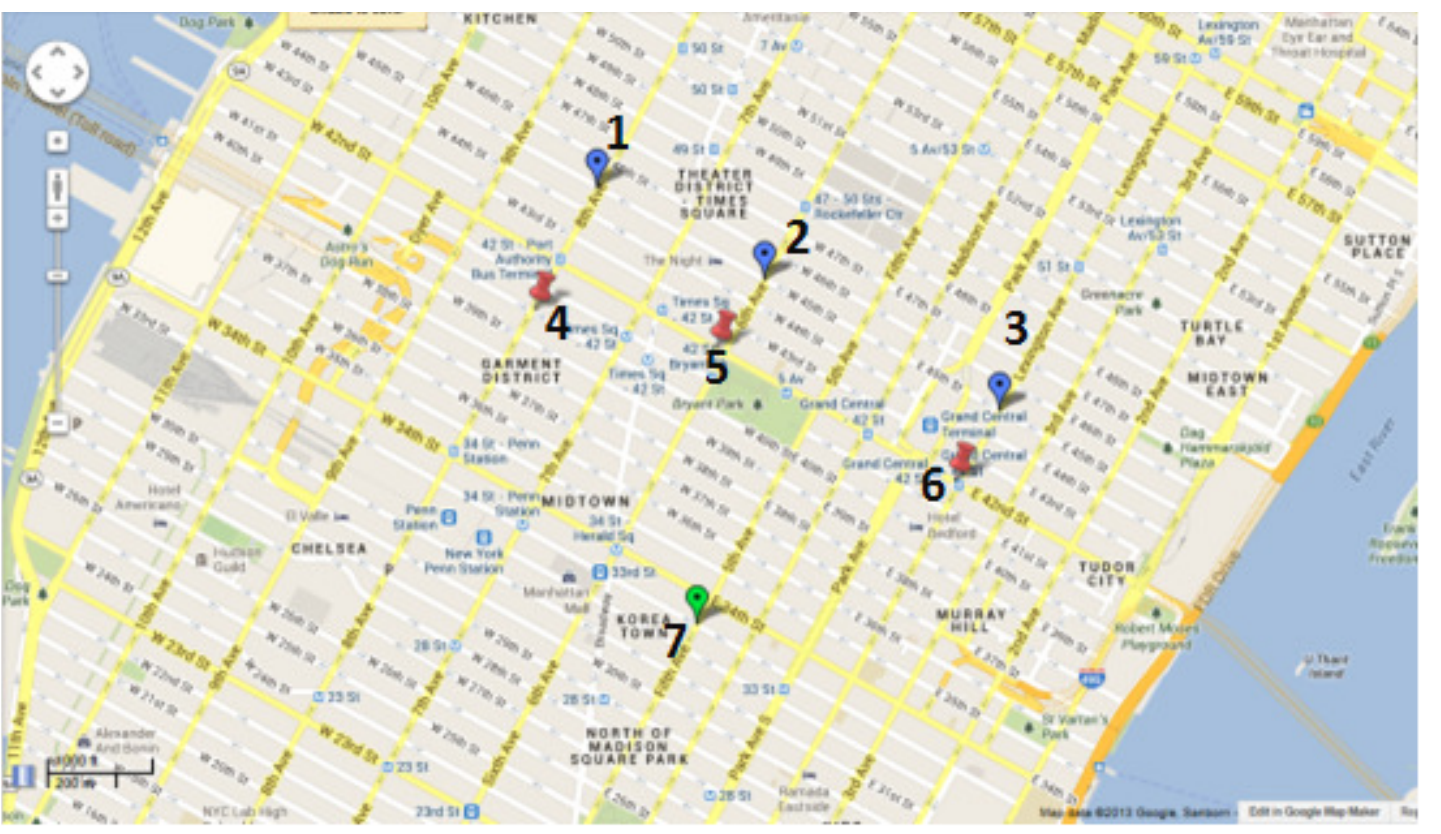}
\caption{An example wireless backhaul network in
downtown Manhattan scenario. Blue, green balloons
and red markers denote edge, gateway and candidate
aggregator node locations respectively.
Our objective is to minimize the aggregator node deployment cost
while ensuring the network connectivity between
edge and gateway nodes. Figure taken from~\cite{GoogleMap}}
\centering
\label{fig:Small_Network}
\end{figure}

Backhaul node placement in urban small cell networks 
differs from the mentioned relay placement problems in 
the following aspects. First, link gain between a candidate aggregator
location and two nearby small cells in a metropolitan setting
can vary significantly because of difference in diffraction angles.
Coverage radius based backhaul node placement becomes
inapplicable. Second, a large subset of relay placement algorithms that assume
an unconstrained number of candidate locations cannot be applied
here since only a subset of building rooftops can be leased.
Third, both sub-$6$ GHz and microwave bands are candidate spectrum
for future generation small cells. The nature of interference pattern and spatial
multiplexing capability varies between the two scenarios and leads
to different network optimization problems. 
Our work encapsulates all these features.
We assume an interference free region and time/frequency division
multiple access while considering microwave band, and 
protocol interference model with space division multiple access 
while considering sub-$6$ GHz band. Using these assumptions,
we optimize the wireless backhaul network in a metropolitan scenario.

The paper is organized in the following way: Section~\ref{sec:Interference_pattern} 
shows interference pattern at different bands. 
Section~\ref{sec:Microwave_Band} and~\ref{sec:Sub6GHz_Band}
show the network optimization problems in microwave band based backhaul
and sub-$6$ GHz band based backhaul respectively. Section~\ref{sec:Optimization_Solution}
presents how we solve the network optimization problems through linear
relaxation techniques and branch-and-bound algorithm. 
After showing the simulation results in section~\ref{sec:Simulation_Results}, we conclude
in section~\ref{sec:Conclusion}.

%The different advantages of different frequencies have led to several different
%choices in wireless networks. First, two hop operation using sub-6 GHz between small cells \& aggregator nodes and millimetre wave band between aggregator nodes \& macrocells;
%second, two hop operation using microwave between small cells \& 
%aggregator nodes and millimetre wave band between aggregator nodes \& macrocells ;
%third, multi-hop daisy chain operation among small cells using millimetre wave band.
%This paper provides the backhaul network optimization
%formulation with these three scenarios.
%
%One needs to lease roof tops of nearby tall buildings to use these as aggregator nodes.
%In a downtown metropolitan city, this lease can consume significant operational expenses.
%Besides, some small cells may not have any nearby tall buildings that are allowed to be leased. On the other hand, some metropolitan small cells may not have any adjacent streets that are allowed to be digged for fiber. 

\section{Interference Models}  \label{sec:Interference_pattern}

Throughout this work, we denote small cells by edge nodes (EN) and
backhaul nodes by aggregator nodes (AN). 
We assume that aggregator nodes
communicate with gateway nodes in millimeter band
using LOS path. Edge nodes can connect with aggregator or gateway nodes in 
microwave band or sub-$6$ GHz band using NLOS path.
We use $5.8$ GHz band, $28$ GHz band and $60$ GHz
as representatives of sub-$6$ GHz, microwave and millimeter wave band
respectively.

%We briefly describe the difference of interference characteristics
%between sub-$6$ GHz and microwave band in the next subsection. Thereafter,
%we show the interference pattern in millimeter wave band.
%we describe the network optimization problem of each of these cases in the
%subsequent subsections. 

\subsection{Interference limited versus interference free setting 
between edge and aggregator/gateway nodes}

Typically, the antennas that operate at $28$ GHz have high gain
and very narrow beam width. We assume that aggregator nodes
perform switched beams and cannot communicate with multiple edge nodes at the same 
time slot. On the other hand, due to the narrow beam width at
both transmitter and receiver antennas, substantial interference suppression
is achieved between non-adjacent links, i.e., two links that do not share a common node. 
We consider time/frequency division multiple access and
interference free regime while considering microwave band between edge nodes
and aggregator nodes.
%Simulation results will justify our interference-free assumption in microwave band.

%The following two sections 

The antennas that operate at sub-$6$ GHz typically come with wide beam width.
An aggregator node can communicate with
different small cells simultaneously using
space division multiple access (SDMA) techniques. However, edge nodes that intend
to communicate with a particular aggregator node
generate interference to neighbouring aggregator nodes.
We consider the spatial multiplexing capability of aggregator nodes 
and use a power control based protocol interference model to 
capture the interference pattern in sub-$6$ GHz band.

\subsection{Interference free setting between aggregator and gateway nodes}

We assume that aggregator nodes - located at the roof tops of tall 
buildings - get LOS paths to gateway nodes and can use millimeter band
for communications to/from the gateway nodes. 
Typically, the antennas that operate at millimeter wave 
band have narrow beam width. We assume that non-adjacent links do
not interfere with each other and nodes cannot perform space division
multiple access due to the complexity of multi-beam
operation. Hence, similar to the microwave band, we assume
an interference free setting and time/frequency division multiple
access based fractional resource allocation in the links that connect
aggregator and gateway nodes.

The next two sections provide the network optimization formulations 
in the following two scenarios: first, edge nodes communicating to
aggregator or gateway nodes using microwave band and second, edge
nodes communicating to aggregator or gateway nodes using sub-$6$ GHz band.
%In both these scenarios, aggregator nodes connect to gateway nodes 
%using interference free millimeter wave band.

\begin{table}
\begin{center}
\begin{tabular}{|l|l|} \hline
Notation & Description \\  \hline
$\mathcal{N}$ & Set of all nodes \\ \hline
$\mathcal{EN}$ & Set of edge nodes \\ \hline
$\mathcal{AN}$ & Set of aggregator nodes \\ \hline
$\mathcal{GN}$ & Set of gateway nodes \\ \hline
$y_j$ & Binary decision variable for AN deployment \\ \hline
$c_{y_j} $ & Operational expense of deployment at node $j$ \\ \hline
%$N$ & Total number of nodes \\ \hline
%$\mathcal{L}$ & Set of sessions \\ \hline
%$L$ & Number of sessions \\ \hline
$\mathcal{M}_{5.8}$ & Set of channels at $5.8$ GHz\\ \hline
$M$ & Number of channels \\ \hline
$f_{ij}$ & Flow between channel $i$ and $j$ \\ \hline
$d_i$ & Demand of edge node $i$ \\ \hline
$W$ & Bandwidth of each channel \\ \hline
$N_0$ & Noise spectral density \\ \hline
$p_{ij}^m$ & Allotted power between node $i$ and $j$ in channel $m$ \\ \hline
$g_{ij}^m$ & Link gain between node $i$ and $j$ in channel $m$ \\ \hline
$x_{ij}^m$ & If node $i$ and $j$ communicate in channel $m$ \\ \hline
$x_j^m$ & If node $j$ uses channel $m$ \\ \hline
$x^m$ & If channel $m$ is used \\ \hline
$T_j$ & Maximum number of radios at node $j$ \\ \hline
$C_{p_i,W_j}$ & Capacity of link evaluated at power $p_i$ and bandwidth $W_j$ \\ \hline
\end{tabular}
\end{center}
\caption{List of Notations} \label{tab:Notations}
\end{table}

%\vspace{-16mm}

\section{Microwave band in NLOS paths - Interference 
free network optimization} \label{sec:Microwave_Band}

We consider a two-hop network with $\mathcal{EN}$ set of edge nodes and  
$\mathcal{GN}$ set of gateway nodes.
Edge nodes act as sources (sinks) and gateway nodes act as sinks (sources)
of data traffic in the uplink (downlink).
Let $\mathcal{AN}$ denote the set of possible node locations
of aggregator nodes.
Aggregator nodes just
relay data between sources and sinks. 
Fig.~\ref{fig:Small_Network} shows an example wireless backhaul
network in downtown Manhattan.

Let us focus on the uplink of a backhaul network.
Assume that $d_i$ denotes the demand of each small cell.
Let $W_{ij}$, $p_{ij}$ and $f_{ij}$ denote the
allotted bandwidth, power and flow of link
$ij$ respectively. Let $y_j$ denote a binary decision variable at node $j$, i.e.,
it represents whether one should place an aggregator node
at the candidate location $j$.
Let $W_{max,b}$ and $p_{max,b}$ denote
the maximum allowed bandwidth and power \emph{per radio} in band $b$.
Node $j$ can deploy up to $T_j$ number of radios.
%In a metropolitan setting, the form factors constraints for
%radio and antenna deployment will vary between the locations
%of edge, aggregator and gateway nodes.
%Hence, the maximum allowed radio per node differs between these
%three types of nodes. 
%We assume $T_j$ to be known
%in advance for each radio node $j$.
Table~\ref{tab:Notations} summarizes the list of notations.
%that we have used throughout the paper.

\begin{figure}[t]
%\small

\begin{subequations}

\begin{equation}
\min \sum_{j \in \mathcal{AN}} c_{y_j} y_j 
\label{eq:Objective1}
\end{equation}
\begin{equation}
\sum_{j \in \mathcal{AN}} f_{ij} + \sum_{k \in \mathcal{GN}} f_{ij}  = d_i 
\, \, \, \, \, \,\forall  i \in \mathcal{EN} 
\label{eq:ENFlow1}
\end{equation}
\begin{equation}
\sum_{i \in \mathcal{EN}} f_{ij} = \sum_{k \in \mathcal{GN}} f_{jk} 
\, \, \, \, \, \,\forall  j \in \mathcal{AN} 
\label{eq:ANFlow1}
\end{equation}
%
%%
%\begin{equation}
%\sum_{k \in \mathcal{GN}} 
%\bigl(\sum_{i \in \mathcal{EN}} f_{ik} + \sum_{j \in \mathcal{AN}} f_{jk} \bigr)
%= \sum_{i \in \mathcal{EN}} d_i 
%\end{equation}
%%
%
\begin{equation}
f_{ij} \leq W_{ij} \log_2 \bigl(1 + \frac{p_{ij} g_{ij,28}}{N_0 W_{ij}} \bigr)
\, \, \forall  i \in \mathcal{EN} \, , \, \forall j \in \mathcal{AN,GN}
\label{eq:ENFlowCapacity1}
\end{equation}
\begin{equation}
f_{ij} \leq W_{ij} \log_2 \bigl(1 + \frac{p_{ij} g_{ij,60}}{N_0 W_{ij}} \bigr)
\, \, \forall  i \in \mathcal{AN} \, , \, \forall j \in \mathcal{GN}
\label{eq:ANFlowCapacity1}
\end{equation}
\begin{equation}
\sum_{j \in \mathcal{AN,GN}} W_{ij} \leq W_{max,28} \, ,
\, \sum_{j \in \mathcal{AN,GN}} p_{ij} \leq p_{max,28} 
\, \, \forall i \in \mathcal{EN}
\label{eq:ENMax1}
\end{equation}
\begin{equation}
\sum_{i \in \mathcal{EN}} W_{ij} \leq y_j \cdot T_j W_{max,28}  
\, \, \forall j \in \mathcal{AN}
\label{eq:ANBWPlacement1}    
\end{equation}
\begin{equation}
\sum_{i \in \mathcal{EN}} W_{ij} \leq T_j \cdot W_{max,28}  
\, \, \forall j \in \mathcal{GN}
\label{eq:GNInBW1}
\end{equation}
\begin{equation}
\sum_{k \in \mathcal{GN}} W_{jk} \leq W_{max,60}  \, \, 
, \sum_{k \in \mathcal{GN}} p_{jk} \leq p_{max,60} 
\, \, \forall j \in \mathcal{AN}
\label{eq:ANMax1}
\end{equation}
\begin{equation}
f_{ij}, W_{ij}, p_{ij} \geq 0 \, \forall \, (i, j) \in \mathcal{E} \, \, ,
y_j \in \{0,1\} \, \, \forall j \in \mathcal{AN}
\label{eq:Var1}
\end{equation}

\end{subequations}
\caption{Network optimization formulation when
edge and aggregator nodes communicate in an interference free setting}
\label{fig:Network_Optimization28}

\end{figure}

Fig.~\ref{fig:Network_Optimization28} shows the network 
optimization formulation of this scenario. Eq.~\eqref{eq:Objective1}
denotes the objective function where we minimize the aggregator node
deployment cost. Eq.~\eqref{eq:ENFlow1} and~\eqref{eq:ANFlow1}
denote the flow conservation constraints. First, each edge
node's outgoing data traffic to the aggregator nodes and gateway
nodes should equal the edge node's demand. Second, each
aggregator node's incoming flow should equal its outgoing flow.
Eq.~\eqref{eq:ENFlowCapacity1} and~\eqref{eq:ANFlowCapacity1}
couple the flow, bandwidth and power variables at each link.
%Commercial $60$ GHz radios require LOS operation and edge
%nodes are unlikely to get LOS path to gateway nodes.
It is assumed that edge nodes use $28$ GHz and aggregator nodes use $60$ GHz. 
Equation~\eqref{eq:ENMax1}-\eqref{eq:ANMax1} denote
the maximum available bandwidth and power constraints at each node.
Equation~\eqref{eq:ANBWPlacement1} couples all other constraints 
with the the aggregator node deployment variable of the optimization objective.
Equation~\eqref{eq:Var1} describes the variables of the optimization
problem.

The optimization problem of Fig.~\ref{fig:Network_Optimization28} 
is a mixed integer non-linear program (MINLP). Next, 
we describe the optimization formulation of sub-$6$ GHz transmission
based networks.
% since it is another MINLP. 
%Thereafter, we provide the solution methodology of these MINLP's.

\section{Sub-$6$ GHz in NLOS paths - 
Protocol interference based network optimization}   \label{sec:Sub6GHz_Band}

Due to the wide beam width of antennas at sub-$6$ GHz,
non-adjacent links can interfere with each other. 
To tackle this interference, 
we split the overall bandwidth at sub-$6$ GHz into a set of discrete channels. 
The edge nodes use these channels to communicate with aggregator or gateway nodes.
We schedule and allocate power in these channels optimally so that non-adjacent links do not interfere with each other.

Let $x_{ij}^m$, $p_{ij}^m$ and $g_{ij}^m$ denote the binary scheduling 
variables, power allocation and gain at link $ij$
in channel $m$ respectively.
\begin{equation}
  x_{ij}^m =\begin{cases}
    1, & \text{if node $i$ transmits to node $j$ using channel $m$}.\\
    0, & \text{otherwise}.
  \end{cases}
\end{equation} 

We use protocol interference model in our work. 
Assume that node $i$ transmits to node $j$
in channel $m$, i.e., $x_{ij}^m = 1$. 
Another node $k$ can transmit to node $h$ in channel $m$ if 
$p_{kh}^m$ causes negligible interference in node $j$.
%%
%\begin{equation}
%x_{ij}^m = 1 \, \implies  \, p_{kh}^m \leq \frac{P_I}{g_{kj}^m}  
%\, \forall (k, h) \in \mathcal{N}, \, k \neq i, \, h \neq j   
%\label{eq:InterferencePower1}
%\end{equation}
%%
%If $x_{ij}^m = 0$, then k's transmission power is bounded by the 
%maximum transmission power, $P_{max}$.
%%
%\begin{equation}
%x_{ij}^m = 0 \,  \implies  \, p_{kh}^m \leq p_{max}  
%\, \forall (k, h) \in \mathcal{N}, \, k \neq i, \, h \neq j   
%\label{eq:InterferencePower2}
%\end{equation}
%%
%Equation~\eqref{eq:InterferencePower1} and~\eqref{eq:InterferencePower2}
%can be combined to the following:
%%
\begin{equation}
p_{kh}^m + (p_{max} - \frac{P_I}{g_{kj}^m}) x_{ij}^m \leq p_{max}  
\, \forall \, k \in \mathcal{N}, \, h \in \mathcal{N}, \, k \neq h  \label{eq:InterferenceConstraint}
\end{equation}
where $P_I$ is the interference threshold.

%All nodes (edge, aggregator and gateway) of the network
%are static. The channel coherence time in a fixed wireless network
%is order of magnitude higher than wireless access.
Due to the wide beam width of sub-$6$GHz antennas,
an aggregator or gateway nodes can cover multiple edge nodes 
using SDMA technology. Hence,
\begin{equation}
\sum_{i \in \mathcal{EN}} x_{ij}^m \leq A \, , 
\, \forall m \in \mathcal{M}_{5.8} \, , \, \forall j \in \mathcal{AN,GN}
\end{equation}
where $\mathcal{M}_{5.8}$ is the set of discrete channels at
$5.8$ GHz and $A$ is the maximum number of edge nodes that one
radio of aggregator/gateway node can cover
using SDMA technology. For simplicity, we assume that the aggregator or
gateway node can employ very large number of antennas at their end
and fully suppress the interference among covered
edge nodes by using a minimum-mean-squared-error decoder
when the ratio of number of antennas to
number of edge nodes becomes very high~\cite{LargeMIMO2}. 
%The value of $A$ depends on the the number of antennas
%at the aggregator/gateway node.
Our model can accommodate the case of imperfect
interference suppression as a gap to capacity.
%The number of covered edge nodes have to 
%less than or equal to the number of antennas of the receiver
%for high interference cancellation among the transmitter nodes.
%Hence, we use $A$ to represent the number of edge nodes
%that an antenna can cover in $5.8$ GHz. 

\begin{figure}[t]
\begin{subequations}

\begin{equation}
\min \sum_{j \in \mathcal{AN}} c_{y_j} y_j 
\label{eq:Objective2}
\end{equation}

\hspace{20mm} Equations \eqref{eq:ENFlow1}, \eqref{eq:ANFlow1},
 \eqref{eq:ANMax1}, \eqref{eq:ANFlowCapacity1}.

\begin{equation}
f_{ij} \leq \sum_{m \in \mathcal{M}_{5.8}} 
W \log_2 \bigl(1 + \frac{p_{ij}^m g_{ij}^m}{N_0 W} \bigr)
\, \, \, \, \, \forall i \in \mathcal{EN} \, , \, \forall j \in \mathcal{AN,GN}
\label{eq:FlowCapacity2}
\end{equation}
%
%%%
%\begin{equation}
%f_{jk} \leq W_{jk,60} \log_2 \bigl(1 + \frac{p_{jk,60} g_{jk,60}}{N_0 W_{jk,60}} \bigr)
%\end{equation}
%%
%\begin{equation}
%\sum_{k \in \mathcal{GN}} W_{jk,60} \leq W_{jmax,60} \, , \,
%\sum_{k \in \mathcal{GN}} p_{jk,60} \leq p_{jmax,60}
%\end{equation}
%%
\begin{eqnarray}
& & p_{kh}^m + (p_{max} - \frac{P_I}{g_{kj}}) x_{ij}^m \leq p_{max} \nonumber \\
& &  \forall k \in \mathcal{EN} , \, h \in \mathcal{AN,GN}, \, k \neq i, \, h \neq j
\label{eq:Protocol_Interference}
\end{eqnarray}
\begin{equation}
\sum_{i \in \mathcal{EN}} x_{ij}^m \leq A y_j \, , 
\, \forall m \in \mathcal{M}_{5.8} \, , \, \forall j \in \mathcal{AN}
\label{eq:AN_Antenna}
\end{equation}
\begin{equation}
\sum_{i \in \mathcal{EN}} x_{ij}^m \leq A \, , 
\, \forall m \in \mathcal{M}_{5.8} \, , \, \forall j \in \mathcal{GN}
\label{eq:GN_Antenna}
\end{equation}
\begin{equation}
\sum_{j \in \mathcal{AN,GN}} \sum_{m \in \mathcal{M}_{5.8}} x_{ij}^m \leq T_j  
\, \, \forall i \in \mathcal{EN}
\label{eq:EN_Antenna}
\end{equation}
\begin{equation}
\sum_{m \in \mathcal{M}} x_j^m \leq T_j  \, \, \forall j \in \mathcal{AN,GN}
\label{eq:ANGN_Channel}
\end{equation}

\begin{equation}
p_{ij}^m \leq p_{max} x_{ij}^m   \, \, \, \, \, \forall (i,j) \in \mathcal{E}, \, \forall m \in \mathcal{M}
\label{eq:PowerScheduling}
\end{equation}
\begin{equation}
x_{ij}^m \leq x_j^m   \, \forall i \in \mathcal{EN}, \, j \in \mathcal{AN,GN} , \, m \in \mathcal{M}_{5.8}
\label{eq:NodeSchedule}
\end{equation}
%%
%\begin{equation}
%x_j^m \leq y_j \, \, \, \, \, \, \forall j \in \mathcal{AN} , \, m \in \mathcal{M}
%\end{equation}
%%
\begin{eqnarray}
& & p_{ij}^m, \, f_{ij} \geq 0,  \, 
x_{ij}^m, \, y_j, \, x_j^m \in \{0,1\}, \,  \nonumber \\
& & \forall i \in \mathcal{EN}, \, j \in \mathcal{AN,GN} , \, m \in \mathcal{M}_{5.8}
\label{eq:Var2}
\end{eqnarray}
\begin{equation}
f_{jk}, \, p_{jk,60}, \, W_{jk,60} \geq 0, \, \forall j \in \mathcal{AN}, \, k \in \mathcal{GN}
\label{eq:Var3}
\end{equation}
\end{subequations}
\caption{Network optimization formulation when
edge and aggregator nodes communicate using a protocol interference model}
\label{fig:Network_Optimization5.8}

\end{figure}

Fig.~\ref{fig:Network_Optimization5.8} shows the network optimization
formulation with the protocol interference and spatial
multiplexing constraints. The optimization objective
of~\eqref{eq:Objective2}, flow conservation constraints 
of~\eqref{eq:ENFlow1},~\eqref{eq:ANFlow1} are same as Fig.~\ref{fig:Network_Optimization28}.
Power and bandwidth allocation equations between aggregator and gateway nodes
(equation~\eqref{eq:ANMax1} and~\eqref{eq:ANFlowCapacity1}) re-appear in
Fig.~\ref{fig:Network_Optimization5.8}.

Equation~\eqref{eq:AN_Antenna} couples aggregator node deployment decision 
variables to all other constraints by ensuring that a
candidate location must be selected for deployment
if it uses any channel. Power control based protocol interference model appears
at~\eqref{eq:Protocol_Interference}. 
Spatial multiplexing capability of aggregator and gateway nodes
appear at~\eqref{eq:AN_Antenna} and~\eqref{eq:GN_Antenna}.
Eq.~\eqref{eq:EN_Antenna} shows that the
number of channels that an edge node can use is limited
by the maximum number of allowed radios in that node.
Eq.~\eqref{eq:ANGN_Channel} denotes that an aggregator or
gateway node $j$ can place up to $T_j$ number of radios.
Eq.~\eqref{eq:PowerScheduling} couples the power allocation
and scheduling variables. Eq.~\eqref{eq:NodeSchedule}
couples the link scheduling and node scheduling variables.
Eq.~\eqref{eq:Var2} and~\eqref{eq:Var3}
describe the variables of the optimization program.

The optimization problem of Fig.~\ref{fig:Network_Optimization5.8}
is also a MINLP. 
%The number of integer 
%variables (both scheduling variable $x_{ij}^m$ and aggregator node
%deployment variable $y_j$) in this scenario is much higher than
%that of Fig.~\ref{fig:Network_Optimization28} which only
%contains deployment variable $y_j$. 
Section~\ref{sec:Optimization_Solution} shows how we solve these MINLP's.

\section{Solution of the Optimization Problem}  \label{sec:Optimization_Solution}

We convert the MINLP's to mixed integer linear programs (MILP)
to speed up the optimization convergence
and to be able to use free solvers.
We relax the log functions of the capacity equations into a set of linear functions, 
solve the resultant MILP using branch-and-bound algorithm 
and find a feasible solution of the optimization problem
from the relaxed solution. We describe these steps in the
next three sub-sections.

\subsection{Linear relaxation of the capacity function}

%%
%\begin{figure}[t]
%\centering
%\includegraphics[scale=0.5]{LogLinearRelaxation.pdf}
%\caption{Linear relaxation of the capacity function.
%First order Taylor approximation at points 
%$p_1$,$p_i$ and $p_{max}$ provide an upper bound of the
%log function.}
%\centering
%\label{fig:LogLinearRelaxation}
%\end{figure}
%%

The capacity functions of~\eqref{eq:ENFlowCapacity1},~\eqref{eq:ANFlowCapacity1}
and~\eqref{eq:FlowCapacity2} are concave functions with respect to 
the allotted power $p$ and bandwidth $W$~\cite{Boyd}. 
%(Concavity of $\log_2 (1+p)$
%can be easily derived by taking double derivative and 
%$W \log_2 (1 + \frac{p}{W})$ is a perspective function 
%with respect to the variable $W$~\cite{Boyd}). 
Hence, each 
capacity function can be upper bounded into a set of linear 
functions by taking slopes at different points~\cite{Kompella}.
Let us define a set of power variables
$\mathcal{P}_I = \{p_1,\cdot,p_i,\cdot,p_{max}\}$
and bandwidth variables $\mathcal{W}_J = \{W_1,\cdot,W_j,\cdot,W_{max}\} $
for a link with gain $g$. 
Let $C = W \log_2(1 + \frac{p g}{N_0 W})$ denote the capacity function
and $C_{p_i,W_j} = W_j \log_2 (1 + \frac{p_i g}{W_j})$
represent the capacity with power $p_i$ and bandwidth $W_j$.
%Now, $\{(p_1,W_1),\cdots,(p_1,W_{max}),\cdots,(p_i,W_j),\cdots,(p_{max},W_1), \cdots,
%(p_{max},W_{max})$ denote the set of power-bandwidth ordered pairs.
%We take derivative at each of these points. Thereafter, we bound
%the flow of the link by each of the linear functions through
%the following function:
We bound the flow $f$ in the link by taking first order Taylor 
approximation in each of the power-bandwidth pairs:
\begin{eqnarray}
f & \leq & C_{p_i,W_j} + m_{p_i} \cdot (p-p_i) + m_{W_j} (W-W_j) \nonumber \\
& & \forall p_i \in \mathcal{P}_I \, , \, \forall W_j \in \mathcal{W}_J
\end{eqnarray}
where,
\begin{equation}
m_{p_i} = \frac{\partial C}{\partial p}\bigg|_{p=p_i} \, , \, 
 m_{W_j} = \frac{\partial C}{\partial W}\bigg|_{W=W_j} 
\end{equation}
%% Important
%%
%\begin{eqnarray}
%& & f \leq c \bigg|_{p=p_i,W=W_i} + \bigl[\frac{\partial c}{\partial p}\bigg|_{p=p_i} ,
%\frac{\partial c}{\partial W}\bigg|_{W=W_j} \bigr] \bigl[(p-p_i), (c-c_j) \bigr]^T \nonumber \\
%& & \, \, \forall p_i \in \mathcal{P}_I \, , \, \forall W_j \in \mathcal{W}_J.
%\end{eqnarray}
%%
%Here, $\frac{\partial c}{\partial p}\bigg|_{p=p_i}$ and $\frac{\partial c}{\partial W}\bigg|_{W=W_j}$
%denote the partial derivatives of the capacity $c$ with respect to the
%power and bandwidth variables, evaluated at $p=p_i$ and $w=w_j$ respectively.
%The capacity of the link with $p_i$ power and $W_j$ bandwidth
%is represented by $c_{ij}$.

% Fig.~\ref{fig:LogLinearRelaxation} shows our linear relaxation procedure.
We relax the non-linear equations of~\eqref{eq:ENFlowCapacity1},~\eqref{eq:ANFlowCapacity1}
and~\eqref{eq:FlowCapacity2} in this way and the MINLP's of Fig.~\ref{fig:Network_Optimization28}
and~\ref{fig:Network_Optimization5.8} convert to MILP's.

\begin{table}[t]
\begin{center}

\begin{tabular}{|r|l|} 
\hline
Line & Operation \\ \hline
$1$ & Assume each edge node (EN) uses its maximum power \\
&  and bandwidth. \\  \hline
$2$ & Calculate the capacity between each edge node to all \\
& aggregator node(AN) and gateway nodes (GN). \\ \hline
$3$ & A link between between EN and AN/GN exists only \\ 
& if it can sustain the demand of the EN. \\ \hline
$4$ & Find the coverage of each AN and GN. \\ \hline
$5$ & Select the GN with the maximum coverage.  \\ \hline
$6$ & Assign all adjacent EN's to this GN. \\ \hline
$7$ & Remove the selected EN's and GN from available set. \\
& Go to line $5$. Iterate until all GN are selected \\
& or all EN's are covered. \\ \hline
$8$ & If all EN's are covered, stop. Else, proceed. \\ \hline
$9$ & Select the AN with the maximum coverage.  \\ \hline
$10$ & Assign all adjacent EN's to this AN. \\ \hline
$11$ & Remove the selected EN's and AN from available set. \\
& Go to line $9$. Iterate until all EN's are covered  \\ \hline
\end{tabular}
\end{center}
\caption{Polynomial Time Aggregator Node Placement Algorithm} \label{tab:PolyTime}
\end{table}

%\vspace{-16mm}

\subsection{Branch-and-bound algorithm}

We use YALMIP~\cite{YALMIP} and GNU Linear Programming Kit (GLPK)~\cite{GLPK} to solve
the MILP's. GLPK uses branch-and-bound algorithm
to solve the MILP. Branch-and-bound algorithm branches in each
binary variable. In each branch, the algorithm calculates a lower bound using continuous relaxation of the 
binary variables and an upper bound by finding a feasible
solution. The algorithm updates the global lower and upper bound
and stops when their difference becomes smaller than the pre-defined optimality gap~\cite{Bertsimas}.

%\subsubsection{Partitioning Approach}

Branch-and-bound algorithm suffers from exponential
worst-case complexity. We select a partitioning approach
to speed up the convergence of the branch-and-bound algorithm. 
We find that aggregator node placement
decision variables ($y_j$) are more important than scheduling
variables $(x_{ij}^m)$. Hence, aggregator node placement decision 
variables are branched before scheduling variables.
Using GLPK~\cite{GLPK} and branch-and-bound method, we can solve an MILP,
consisting of roughly $1000$ binary variables, in $30$ minutes
with $0.5$ optimality gap. We can accept this time complexity
since aggregator node deployment is an offline planning task.

%We should first decide if a candidate
%location is selected for aggregator node deployment and then decide the
%scheduling variables $x_{ij}^m$.

%In standard branch-and-bound procedure, partitioning 
%is done by choosing the variable with the largest relaxation error~\cite{Kompella}.
%The reason for this partitioning is that such variable leads
%to a large gap between the upper and lower bound.
%One should partition this variable such that the relaxation error
%becomes smaller.
%Such standard procedures do not exploit the specific properties
%of the problem. 

%This partitioning speeds up the solution time of our examples. 

\subsection{Feasible solution}

The feasible solution of MILP may not be a feasible solution of the original MINLP
since we relaxed the capacity function into a set of linear functions.
Some edge nodes' flow may exceed the
capacity of their links with the allotted power and bandwidth.
We find a feasible solution in the following ways.

\begin{itemize}

\item \emph{Tightening the relaxation gap:} We increase the 
granularity of piecewise linear approximation.

%we take derivative of the capacity function at a large number of points. 
%This tightens the gap between log function and its linear relaxation,
%which in turn, tightens the gap between the MILP solution and the
%optimal MINLP solution.

\item \emph{Checking for spare bandwidth:} We ensure that 
each aggregator node uses its entire allocated bandwidth
before declaring infeasibility. We find this by fixing
the scheduling and deployment variables of the MILP output,
and running the MINLP for bandwidth, power and flow variables
which is a convex optimization problem.

\item \emph{Iterate the process:} If previous step
does not provide a feasible solution, we iterate the whole process 
by re-formulating the MINLP
where the currently infeasible edge nodes form the new set of edge nodes
and unselected aggregator nodes form the new set of aggregator nodes.

\end{itemize}

\subsection{Special Case: Greedy Set Covering based Aggregator Node Placement}

We can obtain a feasible solution of the optimization problem of Fig.~\ref{fig:Network_Connectivity_5.8}
in polynomial time with the following assumptions:
first, there is no limitation on the number of available discrete channels and
second, an edge node can only talk to one aggregator or gateway node.
A greedy weighted set covering algorithm
can solve this problem. We summarize the algorithm briefly in 
Table~\ref{tab:PolyTime} assuming 
equal deployment cost among all candidate locations.
We skip the details due to lack of space.
Our future work will extend this algorithm to the scenario where the
number of channels is limited.

\section{Numerical results}  \label{sec:Simulation_Results}

\begin{table}[t]
\begin{center}
\begin{tabular}{|c|c|c|c|} \hline
Features & $5.8$ GHz & $28$ GHz & $60$ GHz \\ \hline
Rain Attenuation (dB)~\cite{E-band} & $0$ & $2.5$ & $10$ \\ \hline
Oxygen Absorption (dB)~\cite{E-band} & $0$ & $0.5$ & $15$ \\ \hline
Antenna gain (dB) & $17$~\cite{Ericsson1} & $38$~\cite{Ericsson1} & $38$~\cite{Liberator} \\ \hline
Maximum transmit power (dBm) & $19$~\cite{Ericsson1} & $19$~\cite{Ericsson1} & $25$  \\ \hline
Fading margin (dB) & $15$ & $25$ & $25$ \\ \hline
Channel width (MHz) & $40$~\cite{Ericsson1} & $56$~\cite{Ericsson1} & $160$~\cite{Liberator} \\ \hline
Number of channels & $6$ & $6$ & $6$ \\ \hline
\end{tabular}
\end{center}
\caption{Backhaul features at different bands} \label{tab:Features}
\end{table}

%\vspace{-16mm}

%Fig.~\ref{fig:Original_Network} shows a realistic network
%scenario of downtown Manhattan. 
%We get the channel gain
%at different links of the network in $5.8$, $28$ and $60$ GHz
%bands using ray tracing. 
%We incorporate the backhaul features of table~\ref{tab:Features}
%into the link gain to model the capacity formulations. 
Channel gains, obtained using ray tracing, are taken with the
backhaul features of Table~\ref{tab:Features} to obtain
link capacities.
We assume equal deployment cost for all aggregator nodes'
locations and $100$ Mbps demand from all edge nodes. 

%%
%\begin{figure}[t]
%\centering
%\includegraphics[scale=0.5]{Original_Network.pdf}
%\caption{A realistic network scenario in downtown Manhattan.
%Orange and light blue circles denote locations of 
%edge nodes and gateway nodes. Green circles denote possible
%locations of aggregator nodes.}
%\centering
%\label{fig:Original_Network}
%\end{figure}
%%

\subsection{Network connectivity with microwave band}

\begin{figure}[t]
\centering
\includegraphics[scale=0.5]{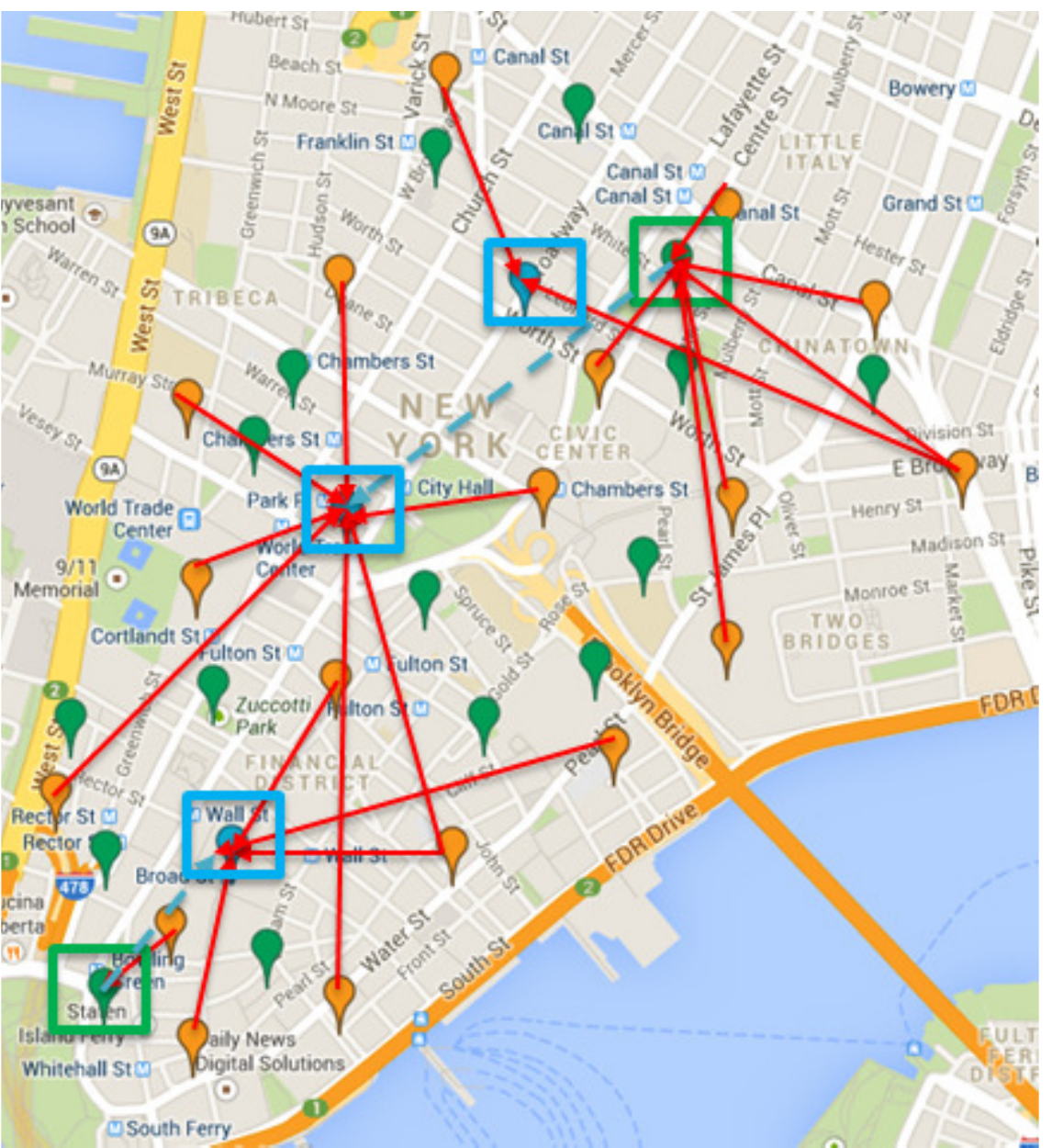}
\caption{Network connectivity when edge nodes (orange markers)
transmit in $28$ GHz (red lines) to aggregator nodes (green markers)
and gateway nodes (highlighted with light blue
rectangle around it). Aggregator nodes transmit in 
$60$ GHz (dashed blue lines) to gateway nodes. Figure taken from~\cite{GoogleMap}}
%and aggregator nodes 
%transmit in $60$ GHz. Orange, light blue and green circles denote  
%edge nodes, gateway nodes respectively and canidate locations
%of aggregator nodes respectively. 
%The figure Network connectivity when edge nodes  
%transmit in $28$ GHz and aggregator nodes 
%transmit in $60$ GHz. Dashed (blue) lines represent
%$60$ GHz based communication and solid red lines represent $28$ GHz based
%communication.}
\centering
\label{fig:Network_Connectivity_28}
\end{figure}

At first, we use the $28$ GHz link gains between the edge and 
aggregator/gateway nodes and $60$ GHz link gains 
between aggregator and gateway node. We run the network
optimization problem of Fig.~\ref{fig:Network_Optimization28}.
Fig.~\ref{fig:Network_Connectivity_28} shows the network
connectivity in this scenario. Two candidate aggregator locations 
-- highlighted with green rectangle marker around them --
get selected for aggregator node deployment. 
The optimality gap is $0\%$.
%Several edge nodes transmit directly to the gateway node.
%Since aggregator node deployment consumes additional cost,
%An edge node routes its traffic through aggregator node
%only when it does not find a good path to any of the gateway nodes.

Our resultant network is free of primary interference.
Adjacent links use different bandwidth in the network
scenario of Fig.~\ref{fig:Network_Connectivity_28}. 
However, non-adjacent nearby links are allowed to share bandwidth. 
This happens since we assumed an interference free regime 
in the network optimization formulations of microwave band.
We now check the validity of our assumptions. 
Assuming that antennas have no side lobes,
we find that the maximum interference among non-adjacent links
that use microwave band fall 19 dB below the noise threshold.
%Hence, our assumption of interference free regime
%is justified in the microwave band.

%

\subsection{Network connectivity with sub-$6$ GHz band}

In this setup, we use the $5.8$ GHz link gains between edge and aggregator/gateway
nodes and $60$ GHz links gains between aggregator and gateway nodes.
Using these link gains, we run the optimization problem of Fig.~\ref{fig:Network_Optimization5.8}. 
We assume an aggregator/gateway node
can cover up to four edge nodes in the same channel
using SDMA at $5.8$ GHz. 

\begin{figure}[t]
\centering
\includegraphics[scale=0.5]{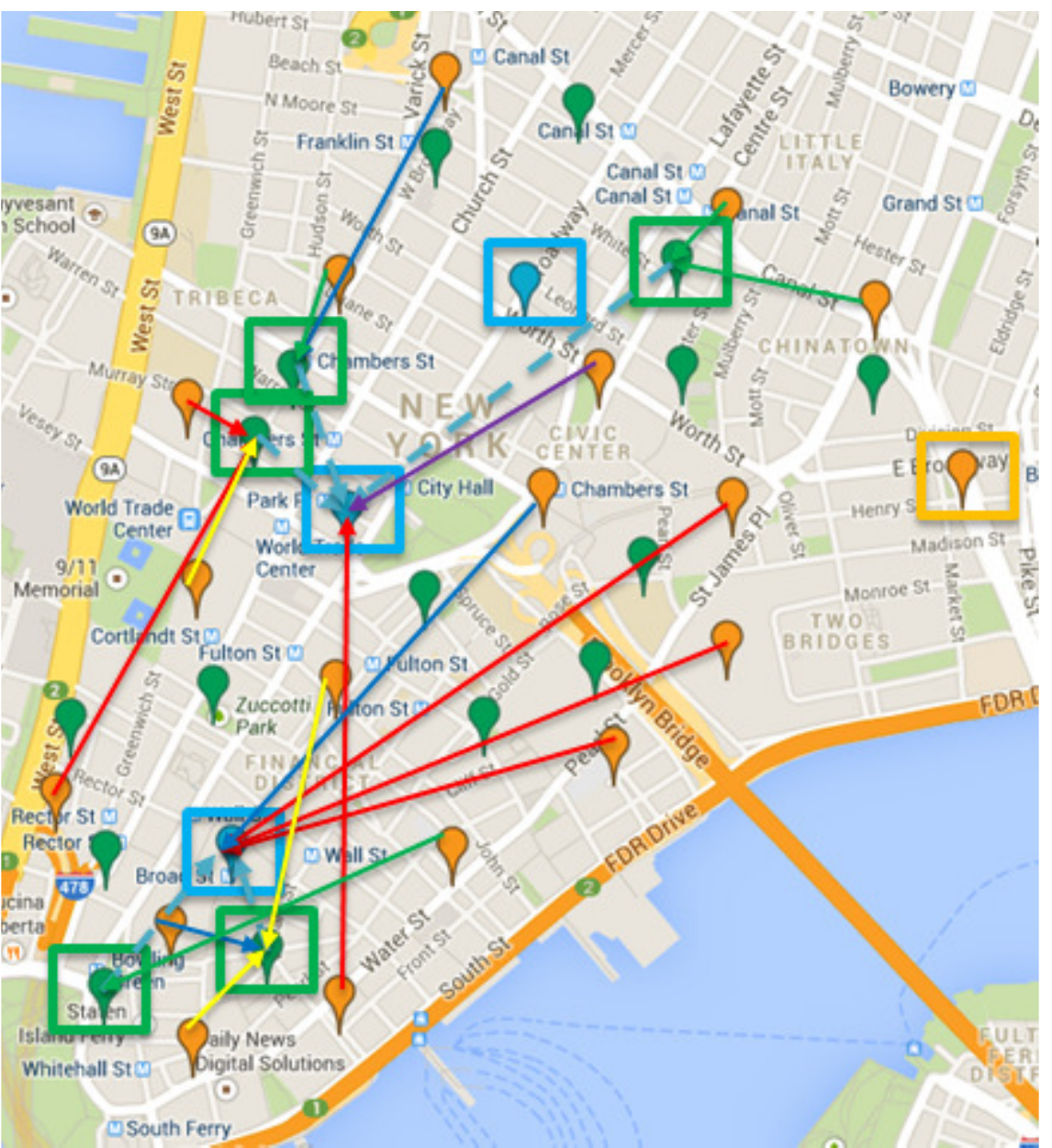}
\caption{Network connectivity when edge nodes 
transmit in $5.8$ GHz and aggregator nodes 
transmit in $60$ GHz. Figure taken from~\cite{GoogleMap}}
\centering
\label{fig:Network_Connectivity_5.8}
\end{figure}

Fig.~\ref{fig:Network_Connectivity_5.8} shows the associated network connectivity.
Each solid colored line represents a discrete channel of $5.8$ GHz
channel set. Some aggregator/gateway nodes communicate to multiple edge
nodes in the same discrete channel using spatial multiplexing 
capability. Two non-adjacent nearby links perform power
allocation and get colored in such a way so that no edge
interferes with each other. One Edge node (highlighted with orange 
rectangle marker around it) does not have good enough
link gain with any aggregator or gateway node to sustain its demand.
It becomes an infeasible edge node. 
The rest of the edge nodes require the deployment of five aggregator nodes
-- highlighted with green rectangle marker around them --
to meet their demand. The optimality gap is $40\%$ in this case.
We ran MILP of both sub-$6$ GHz and microwave band for $30$ minutes.
Optimization problem of Fig.~\ref{fig:Network_Optimization5.8} contains higher number of
binary variables (both scheduling and node placement variables)
than that of Fig.~\ref{fig:Network_Optimization28} (only node placement variables).
Hence, sub-$6$ GHz based network optimization converges slowly.

We do not model many practical aspects such as antenna alignment, material 
reflectivity, etc. that affect the link gain at $28$ GHz. We do not intend 
to compare sub-$6$ GHz and $28$ GHz band.
We just contrast their respective optimization problems.

%\subsection{Performance comparison between microwave
%band and sub-$6$ GHz}
%
%We assume $38$ dB antenna gain at microwave band and
%$17$ dB antenna gain at $5.8$ GHz band~\cite{Ericsson1}.
%Microwave band suffers from higher path loss (due to high frequency),
%higher fading margin, oxygen and rain absorption loss than sub-$6$ GHz. In 
%spite of accounting for all these factors, $42$ dB ($ = 2 * (38 - 17)$) boost
%that come from the increased antenna gains of transmitter and receiver
%ensure that the final link gains of microwave band remain higher than
%that of sub-$6$ GHz band.
%On the other hand, sub-$6$ GHz band can spatially multiplex up to four edge nodes
%in each discrete channel. We do not assume this spatial multiplexing capability
%in microwave band.
%
%
%Our simulation results show that sub-$6$ GHz band based wireless backhaul network
%consumes more operational expenses than that of microwave
%band based wireless backhaul network in this particular scenario.
%The high antenna gain advantage of microwave band
%outweighs the spatial multiplexing advantage of
%sub-$6$ GHz band due to the low signal-to-noise-ratio (SNR) links 
%in this network. Spatial multiplexing capability of sub-$6$ GHz
%may outweigh the antenna gain of microwave band in a network
%where edge nodes and aggregator/gateway nodes have high SNR links.

\section{Conclusion}  \label{sec:Conclusion}

Small cells can keep up with the increasing demand
of wireless networks; 
but require backhaul to transport data to(from) a gateway node.
Wireless backhaul can provide an inexpensive 
option to small cells. Aggregator nodes,
located at roof tops of tall buildings near small cells,
can provide high data rate to multiple small cells in NLOS paths
, sustain the same data rate to gateway nodes in LOS paths
and take advantage of all available bands for wireless backhaul.

This work performed
joint cost optimal aggregator node placement, power allocation, 
channel scheduling and routing to optimize the wireless backhaul network. 
We investigated wireless backhaul network using 
both sub-$6$ GHz and microwave bands.
We considered the different interference patterns 
and multiple access features in these bands and incorporated
them in backhaul network optimization.
Future works will include SINR based interference methodology
(rather than protocol) and mixed wired/wireless bakchaul optimization.
%Our future work will use signal-to-interference-plus-noise-ratio
%model to capture inter-link interference.

%Our results suggest that the high antenna gain advantage of
%microwave band may outperform the spatial multiplexing capability
%of sub-$6$ GHz band in the non-line-of-sight paths of metropolitan
%scenarios. However, due to the low beam width of microwave antennas,
%both transmitter and receiver antennas need to aligned precisely
%to capture the diffraction angle in a metropolitan scenario.

%In a metropolitan setting, some edge nodes may not have any
%nearby tall buildings whose roof tops can be leased. These edge
%nodes need to be connected through fiber from the gateway node.
%Network optimization with mixed wired-wireless backhaul
%remains an area of our future research.

\bibliographystyle{IEEEbib}
\bibliography{BibCISS}

%% Extra stuff
% Modeling optimization problem along the lines of
% Wei Yu's paper
%\begin{equation}
%\min \sum_{j \in \mathcal{AN}} c_{y_j} y_j 
%\end{equation}
%
%\begin{equation}
%\mathbf{f_{ij}} \in \mathcal{N}(\mathbf{d}) \, \, \forall \, (i, j) \, \in \, \mathcal{E}
%\end{equation}
%
%\begin{equation}
%\mathbf{f_{ij}} \in \mathcal{C} \bigl( \mathbf{W_{ij,28}},
%\mathbf{p_{ij,28}} \bigr) \, \, \forall i \in \mathcal{EN}, \, 
%\forall j \in \mathcal{AN,GN}
%\end{equation}
%
%\begin{equation}
%\mathbf{f_{jk}} \in \mathcal{C} \bigl( \mathbf{W_{jk,60}},
%\mathbf{p_{jk,60}} \bigr) \, \, \forall j \in \mathcal{AN}, \, 
%\forall k \in \mathcal{GN}
%\end{equation}
%
%\begin{equation}
%\mathbf{f_{ij}} \, , \, \mathbf{W_{ij}} \, , \, \mathbf{p_{ij}} \, \geq 0 \, 
%\forall \, (i, j) \, \in \, \mathcal{E} \, , 
%\, \,  y_j \in \{0,1\} \, \forall \, j \, \in \mathcal{AN}
%\end{equation}
%
%\end{subequations}

\end{document}